\documentclass[fleqn,12pt,twoside]{article}
\usepackage{espcrc1}
\usepackage{epsf}
\usepackage{pslatex}

\usepackage{graphicx}
\usepackage[figuresright]{rotating}

\newcommand{\AmS}{{\protect\the\textfont2
  A\kern-.1667em\lower.5ex\hbox{M}\kern-.125emS}}
\newcommand{\be}{\begin{eqnarray}}
\newcommand{\ee}{\end{eqnarray}}
\newcommand{\non}{\nonumber\\}
\newcommand{\raf}[1]{(\ref{#1})}
\newcommand{\ave}[1]{\left\langle #1 \right\rangle}

\title{Some remarks on the statistical model of heavy ion collisions}

\author{V. Koch\address{Nuclear Science Division, Lawrence Berkeley
    National Laboratory, \\ 
    1 Cyclotron Road, \\
   Berkeley, CA 94720, USA}
  }
\begin{document}

\maketitle

\begin{abstract}
This contribution is an attempt to assess what can be learned from the
remarkable success of the statistical model in describing ratios of particle
abundances in ultra-relativistic heavy ion collisions.
\end{abstract}

\section{INTRODUCTION}
As already pointed out in the contribution by A. Bialas \cite{Bialas}
the statistical model (see e.g. \cite{stat_model})
works very well in describing/predicting measured
ratios of particle abundances in ultra-relativistic heavy ion collisions. But
even more remarkably, it works also for particle ratios measured in high
energy proton-proton and even $e^+e^-$ collisions. In this contribution we
will take the success of the statistical model as given and rather ask
ourselves what can be learned from that. A critical discussion of possible
shortcomings of the statistical model is given in the contribution of J. 
Rafelski \cite{Rafelski}.

In general a statistical description of a physical system is appropriate if the
system has many degrees of freedom but is characterized only by few
observables/measurements. This is e.g. the case in a thermal system, which is
characterized only by the constants of motion, namely the energy (and
momentum), volume and all the conserved particle numbers. (Of course in a
canonical or grand canonical formulation, the energy and/or particles number
are replaced by the conjugate variables temperature and chemical potential). 

But not only a thermal system meets the requirements for a statistical
description. Let us consider a high energy collision which produces many
particles in the final state. If we are only interested in the number of pions
produced, we constrain the final state very little, and thus statistical
methods should be applicable. This is the idea of the statistical theory of
particle production first invented by Fermi \cite{Fermi}. 

Now suppose the statistical model of Fermi applies for particle production in
high energy collisions. Does that mean that we are dealing with a thermal
system in the sense of Boltzmann, where particle collisions keep the system in
a state of equilibrium? This is very unlikely in case of $e^+e^-$ collisions,
where the produced particles hardly have a chance to re-interact. And actually
explicit measurements \cite{jets} show no indication for interaction among the 
partons from different jets in $e^+e^-$ collisions (see also contribution by
H. Satz \cite{Satz}). Therefore, ``statistical'' does not always mean 
``thermodynamic'' in
the sense that one is dealing with matter in thermal equilibrium, and that one
can define a pressure and an equation of state. Statistical may simply mean
phase-space dominance and the "temperatures" and "chemical" potentials are
nothing but Lagrange multipliers characterizing the phase-space integral
\cite{Hagedorn,Kajantie,Landau}

This, however, may be different in a heavy ion collision. There one would
naively expect (this is actually the main motivation for such complicated
experiments) that the initially produced particles do re-interact on the
partonic and/or hadronic level. The question then is, how to experimentally 
establish that a sufficient amount of  re-interaction has taken place and that
matter in the Boltzmann sense has been formed.

This contribution is organized as follows. In the first section, we will
discuss the phase-space (or statistical model) 
for elementary collisions such as
$e^+e^-$. Then we will proceed with nucleus-nucleus collisions. Finally we will
try to assess to which extent a case for thermal matter can be made in 
nucleus-nucleus collisions. We will conclude with a discussion on what the
statistical
variables extracted from particle ratios can tell us about the phase structure
of QCD.

\section{PHASE-SPACE DOMINANCE} 
Let us consider a high energy collision of elementary particles such as 
$e^+e^-$ or proton proton. The probability to produce $n$ particles 
of a given species, such as pions, is  given by
\be
P_n = \sum_x P_{n,x}
\ee
where $P_{n,x}$ denotes the probability to find $n$ particles of interest 
and $x$ other particles in the final state,
\be
P_{n,x} \sim \int \prod_{i=1}^m \frac{d^3 q_i}{E_i} 
\left| M(q_1,\ldots q_m;\,p_1,p_2) \right|^2 
\delta (E - ( \sum_{i=1}^m E_i ) \,
\delta^{(3)}(\sum_{i=1}^m \vec{q}_i ).
\label{eq:pn}
\ee
Here $E$ is the total energy of the system, which we consider in the center
of momentum frame. The total multiplicity $m$ is given by 
\be
m = n+x.
\ee 
 
In case of many particles in the final state, $m \gg 1$, one integrates over a
large phase-space volume. As a result the details of the matrix element 
$M(q_1,\ldots,q_m;\,p_1,p_2)$ become less relevant. Instead one is sensitive to
a phase-space average of the matrix element. Thus we can rewrite
eq. \raf{eq:pn} as
\be
P_{n,x} &\sim& 
\left[ \frac{1}{V^m} 
\ave{\frac{ \left| M(q_1,\ldots,q_m;\,p_1,p_2) \right|^2}{\prod_{i=1}^m E_i} }
\right]
\,\,\, V^m\int \left(\prod_{i=1}^m d^3 q_i \right)  
\delta (E - \sum_{i=1}^m E_i) \,
\delta^{(3)}(\sum_{i=1}^m \vec{q}_i ) \non
\non
& = &
\bar{S}_m \Phi_m(E)
\label{eq:pn_ave}
 \ee
 where 
 \be
 \Phi_m(E) =V^m\int \left(\prod_{i=1}^m d^3 q_i \right)  
 \delta (E - \sum_{i=1}^m E_i) \,
 \delta^{(3)}(\sum_{i=1}^m \vec{q}_i ) \non
 \ee
 is the micro-canonical m-particle 
 phase-space volume known from statistical physics, and
 \be
 \bar{S}_m &=& \left[ \frac{1}{V^m} 
 \ave{\frac{ \left| M(q_1,\ldots,q_m;\,p_1,p_2) \right|^2}{\prod_{i=1}^m E_i} }
 \right] \non 
 & =& \frac{1}{\Phi_m(E)} \int \prod_{i=1}^m \frac{d^3 q_i}{E_i} 
 \left| M(q_1,\ldots,q_m;\,p_1,p_2) \right|^2 
 \delta (E - \sum_{i=1}^m E_i) \,
 \delta^{(3)}(\sum_{i=1}^m \vec{q}_i )
\label{eq:sn}
\ee
denotes the phase-space averaged m-particle matrix element.
 
Obviously, if $\bar{S}_m$ 
is simply a constant, independent of $m$, and thus independent on $n$ and $x$,
the relative probability to find a given number of particles is simply 
given by the ratio of the phase-space volumes,
\be
\frac{P_n}{P_{n'}} = \frac{\Phi_n(E)}{\Phi_{n'}(E)},
\ee
or in other words, it is given by statistics only. 

Similarly, the mean number
of particles in this case is, up to  a constant, given by statistics
\be
\ave{N} = \sum_n n P_n \simeq \bar{S} \sum_{n,x}n \, \Phi_{n+x}(E),
\label{eq:n_av}
\ee 
where $\bar{S}$ denotes the constant averaged matrix element. Obviously, in
this case particle ratios are given only by statistics, as the constant
$\bar{S}$ drops out. Note, that for a large average multiplicity
  $\ave{m}=\ave{n+x}$, the sum in eq. \raf{eq:n_av} will be dominated by a
  few terms with $n+x \simeq \ave{m}$. This is analogous to the 
the grand-canonical approximation in statistical physics.  

If the mean multiplicity is large, $\ave{m}   \gg 1$, then the
micro-canonical phase-space volume $\Phi_m(E)$ may be evaluated in the
canonical 
or grand-canonical approximation \cite{Hagedorn,Kajantie,Landau} 
leading to Lagrange multipliers, which in the
thermodynamic framework are the temperature and the chemical potential. In the
situation at hand, however, these Lagrange multipliers do not have a
physical meaning. They simply characterize the phase-space integral. Their 
actual magnitude depends on the
available energy as well as on the density of states, i.e., the hadronic mass
spectrum. They, however, do
not reflect exchange of energy with a heat-bath, as is the case for the
temperature in the canonical ensemble of thermal 
physics. 
Thus, in order to avoid confusion, we will denote the application of
statistical physics in the non thermodynamics context by {\em "phase-space
  dominance"}.

\subsection{Conditions on the matrix elements}
As discussed above, the essential assumption for phase-space dominance to
work is that the phase-space averaged matrix elements \raf{eq:sn} are constant,
independent of $m$. What  requirements does this impose on the 
matrix elements? Obviously, if the matrix elements simply scale with the
multiplicity $m$ like
\be
\left| M_m \right|^2 = C \,\, (V^m \prod_{i=1}^m E_i)
\label{eq:const_M}
\ee
with $C$ being a constant, the condition is fulfilled. The scaling with
$ \prod_{i=1}^m E_i$ is simply due to the normalization of the states, and
thus is not a dynamical constraint. The scaling with $V^m$ on the other hand is
not trivial and implies that there is only one relevant length/mass scale 
in the
problem. Before we discuss this in more detail let us list other conditions,
which the matrix element has to satisfy.

\begin{itemize}
\item Absence of strong correlations. Correlations imply that the
  matrix element provides more support in localized regions of phase
  space. Consequently it is far from being constant. Or in other words, the 
  integral in
  eq. \raf{eq:sn} will only have support in a limited region leading to a
  decrease of $\bar{S}_m$ with increasing $m$.
\item Absence of strong energy dependence in the matrix element. This is
  similar to the previous condition and actually related. Strong energy
  dependence (other than the trivial one from the normalization factors of the
  states) obviously implies a non-constant matrix element.   
\item Absence of strong interference effects, 
  which lead to both correlations and  energy dependencies. 
\end{itemize}

\begin{figure}[htb]
\epsfxsize=0.6 \textwidth
 \centerline{\epsfbox{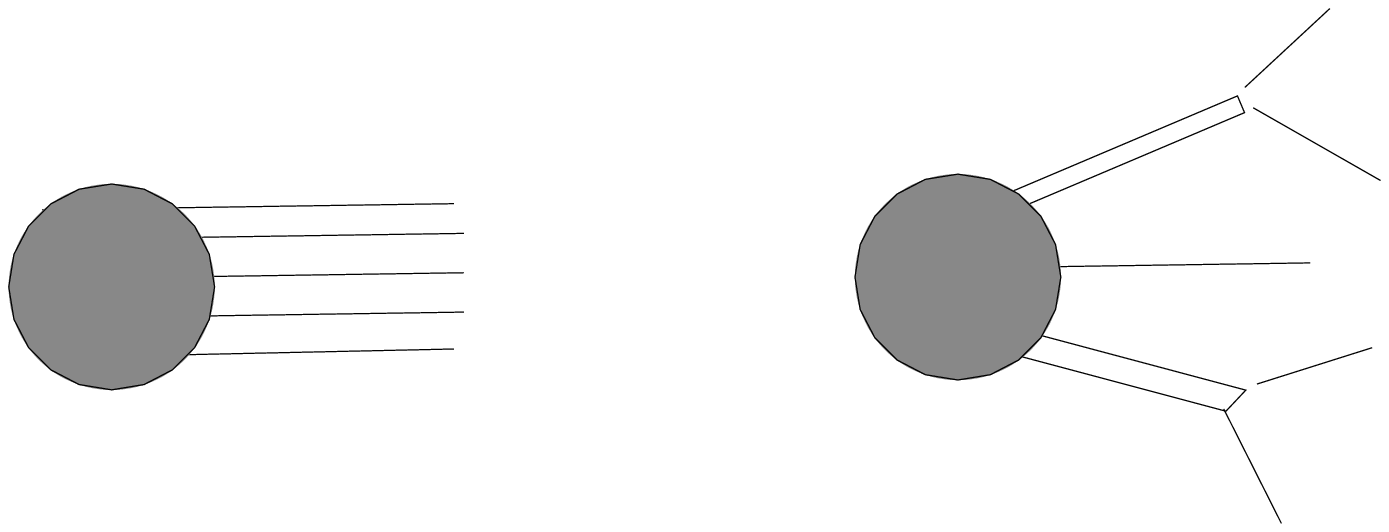}}   
 \caption{Resonances in the final state.}
\label{fig:1}
\end{figure}

Hadronic resonances, such as $\rho$ mesons give rise to correlations and energy
dependencies. And indeed, the statistical model fails to reproduce the data if
only true final state particles such as pions, kaons etc. are taken into
account \cite{becattini_private}. Instead, the successful fits of the particle
ratios are obtained only if the hadronic resonances are part of the
statistical ensemble. This way, the correlations are removed from the matrix
elements and put into the "final" states, in the spirit of
\cite{beth_uhlenbeck}. This is schematically depicted in Fig.\ref{fig:1}.  
Thus, the relevant phase-space to be considered is a phase-space of all
hadronic resonances and the matrix element is reduced to one with
resonances in the final state
\be
\Phi_n(E) & \Rightarrow & \Phi_{n'}^R(E) \non
M(E;q_1,\ldots,q_m) & \Rightarrow & M_R(E;p^R_1,\ldots,p_{m'}^R)
\label{eq:m_red}
\ee
As a result the reduced matrix element $M_R$ is free of all the correlations
introduced by the resonances and, therefore, it is more likely that it meets
the requirements stated above.

Let us return to the issue of the volume dependence of the matrix element.
Only if the matrix element scales with the volume as given
by eq. \raf{eq:const_M}, the statistical approach is justified. From
dimensional arguments an m-particle matrix element has the correct scaling
behavior. 
\be
\frac{\left|M_m\right|^2}{E^m} \sim [{\rm mass} ]^{-(3m + 4)}.
\ee
In general there may be several length/mass scales contributing to 
the matrix element, such as e.g. the hadronic resonances. In this case the
statistical approach should not work. 
If, on the other hand, all the dynamical mass scales of QCD aside from
$\Lambda_{QCD}$  are the masses of the resonances, then the reduced matrix
element $M_R$ \raf{eq:m_red} contains only $\Lambda_{\rm QCD}$, and the
statistical approach will work as long as the volume is of the size of
$1/\Lambda_{\rm QCD}^3$. This would be about the size of the proton, which
appears to be a reasonable size for a volume in an elementary particle
collision. We should note, however, that the fits to proton-proton collisions
\cite{becattini_heinz} lead to volumes of the order of 
$20 \, \rm fm^3$, which is
somewhat on the large side of what one would expect from our considerations
here.  

The constraints on the intrinsic mass 
scales, however, are not as severe as it might
appear from the previous considerations. If the mean multiplicity $\ave{m}$
is large, 
the particle production is dominated by events with final state multiplicities
near $\ave{m}$, i.e. $m = n+x $ is approximately constant for all n. 
And, therefore, the
condition \raf{eq:const_M} is fulfilled trivially.

Finally, the matrix element is responsible for conservation laws due to
intrinsic symmetries, such as strangeness, charge and baryon number. This,
however, is already accommodated in the statistical approach. 
If the amount of conserved quanta is small, one may
have to use a canonical description instead of a grand canonical one. But this
is all within the framework of statistical physics, which actually is based on
conservation laws.

\begin{figure}[htb]
\epsfxsize=0.6 \textwidth
 \centerline{\epsfbox{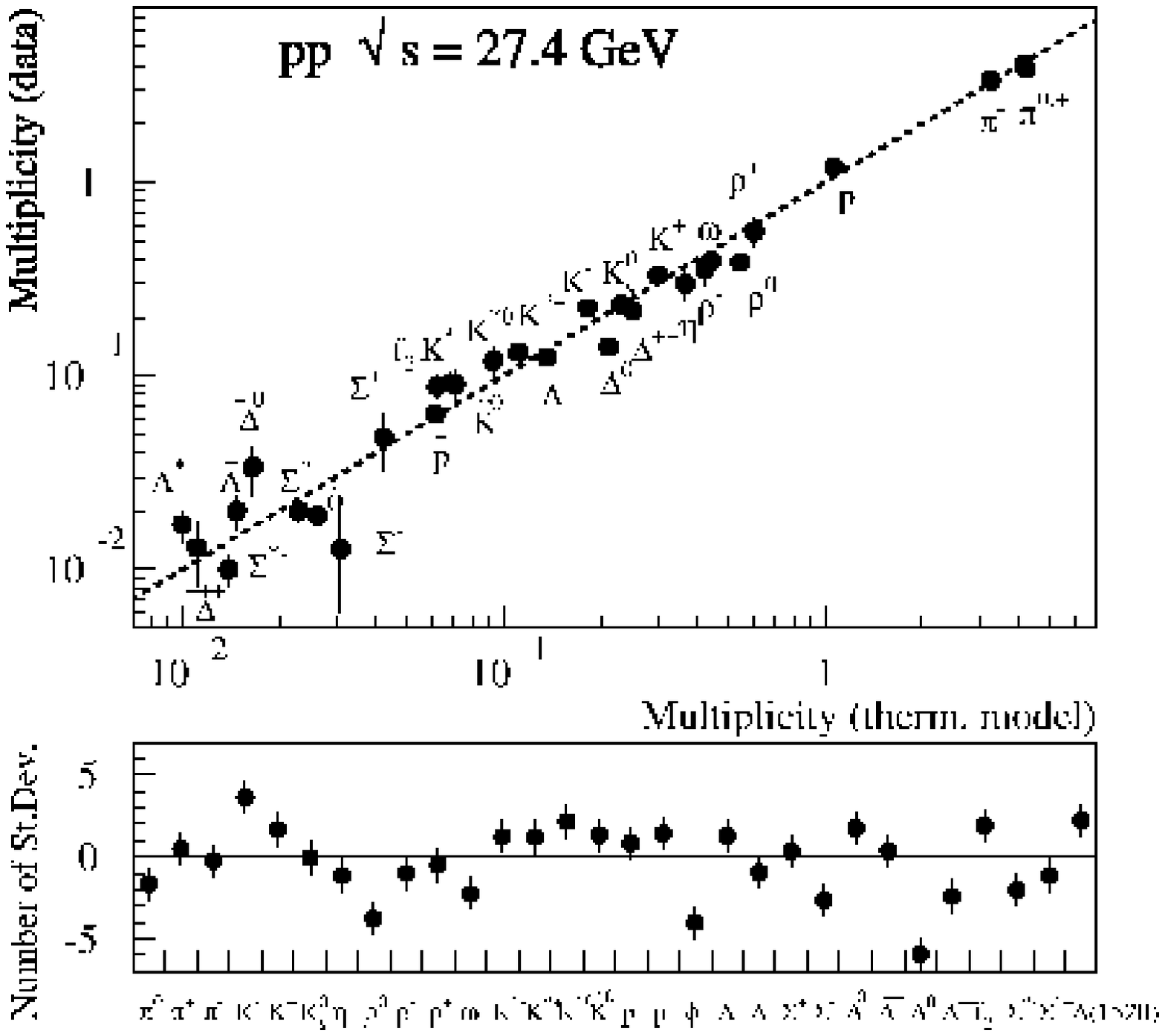}}   
 \caption{Fit of statistical model to particle ratio in proton-proton
 collisions (from \protect\cite{becattini_heinz}).}
\label{fig:2}
\end{figure}

The success of the statistical fits to the particle
ratios for $e^+e^-$ as well as proton proton is demonstrated in 
Fig. \raf{fig:2}. As already pointed out, these fits are based on a
statistical ensemble of hadronic resonances. Following the above arguments, we
may conclude from the success of the statistical model that the
relevant dynamical correlations and mass scales of QCD are contained in 
the hadronic resonances. In order to see more subtle dynamic effects, one
probably has to resort to higher order correlations. The alternative
conclusion would be that even in $e^+e^-$ collisions, re-scattering leads to a
true thermodynamic system. This, however, is difficult to imagine and 
there is no experimental evidence for any re-scattering \cite{jets,Satz}. 

Finally, let us point out that the statistical model also seems to work for
correlation measurements of strange particles \cite{becattini_heinz}, such as
$K_s \,K_s$. In the statistical approach, these correlations are mostly due to
strangeness conservation and the agreement with the data indicates the absence
of strong dynamical correlations.

\section{NUCLEUS-NUCLEUS COLLISIONS}
As we have discussed above, the success of the statistical model in describing
particle yields in proton-proton collisions can be understood as a result of
phase-space dominance. The goal of nucleus nucleus collisions, however, is to
create matter, i.e., a thermal system in the sense of Boltzmann, where particle
collisions lead to and maintain thermal equilibrium. It is only in this
situation, where we can give the Lagrange multipliers "$T$" and "$\mu$" the
physical meaning of temperature and chemical potential.
 
\begin{figure}[htb]
\epsfxsize=0.8 \textwidth
 \centerline{\epsfbox{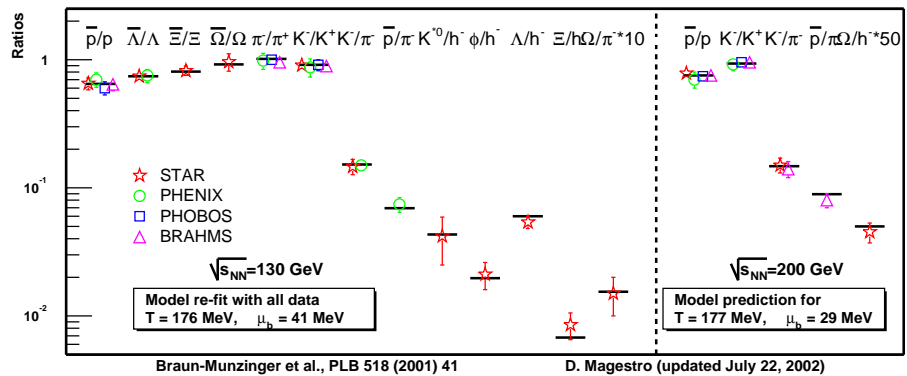}}   
 \caption{Fit of statistical model to particle ratio in Au+Au collisions at
 RHIC energies \protect\cite{pbm2}.}
\label{fig:2b}
\end{figure}

Obviously, if each individual nucleon-nucleon collision can be described by a
statistical approach,  we expect the statistical model to work even better in 
a nucleus-nucleus collision. And indeed it does, as can be seen from
Fig.\ref{fig:2b}.  But how do we know that the statistical behavior of a 
nucleus-nucleus collision is again not simply phase-space dominance?

\begin{figure}[htb]
\epsfxsize=0.6 \textwidth
 \centerline{\epsfbox{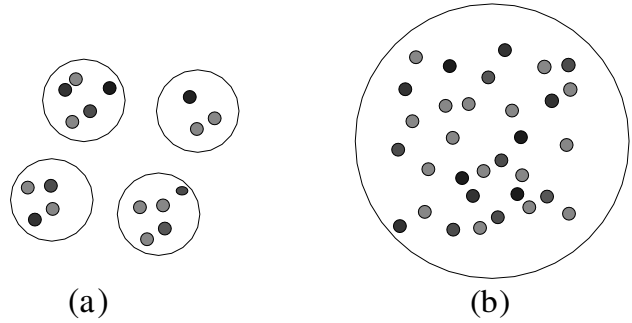}}   
 \caption{Individual nucleon-nucleon collisions (a) and nucleus-nucleus
 collision (b). }
\label{fig:3}
\end{figure}

To illustrate this point, let us assume for a moment that a nucleus-nucleus
collision is a simple superposition of ``$N$''
nucleon-nucleon collisions. Let us also 
assume that nucleon-nucleon collisions can be described by statistics as a
result of phase-space dominance. If we were dealing with a simple classical
ideal gas without additional constraints from conservation laws, the 
partition function of the
nucleus-nucleus system $Z_{AA}$ is simply the product of the partition
functions of the nucleon-nucleon collisions $Z_{nn}$
\be
Z_{AA}^{(a)} = \prod_{i=1}^N Z_{nn}.
\ee
This situation is sketched in Fig.\ref{fig:3}a. There is {\em no} 
cross talk between the individual systems.  
On the other hand $Z_{AA}^{(a)}$ also represents the partition function of a 
system of volume 
\be 
V_{AA} = N V_{nn}
\ee
which would correspond to the system depicted in Fig.\ref{fig:3}b, and thus
to ``matter'', or in other words 
\be
Z_{AA}^{(a)} = Z_{AA}^{(b)}
\label{eq:factorize}
\ee
So in this case there would be simply no way to distinguish
between situation (a) and (b) in Fig.\ref{fig:3} within a statistical
framework.  

So how can we find out if indeed thermalized 
matter has been created in a heavy ion collisions?

Obviously the factorization condition \raf{eq:factorize} will break down, once
we probe the boundaries of phase-space available for a nucleon-nucleon
collision, where the statistical model will not work. This could for example
be achieved by studying $n$-particle correlations, with $n$ larger than the
average multiplicity of a nucleon-nucleon collision, $n \gg \ave{N}_{nn}$. If,
such a n-particle correlation would still look "thermal" in an AA collision,
then the vastly bigger phase space of an $AA$-system has been populated by
scattering processes, and we may talk about ``matter''.

The sensitivity of this approach can be improved by looking at conserved
quantum numbers. If additional conservation laws, such as strangeness, 
are at work, phase space
is even more restricted and factorization may  break down already on the single
particle level. Consider for example
strangeness conservation. In scenario (a), strangeness has to be conserved for
each nucleon-nucleon collision separately, whereas in (b) conservation
applies only to the entire system. This additional constraint is most relevant
if the number of strange particles produced in a nucleon-nucleon collision is
small, $N_s \le 1$. If $N_s \gg 1$ a grand-canonical treatment is adequate and
factorization \raf{eq:factorize} works again at least on the single particle
level. Consequently, in this case multi-particle correlations need to be
investigated.  

In nucleon-nucleon collisions a canonical treatment, where strangeness as well
as baryon number are conserved explicitely, 
is required  to explain the particle
abundances \cite{becattini_heinz,redlich-canonic}. 
Also for lower energy and peripheral heavy ion collisions, 
the explicit treatment of strangeness conservation seems to required 
\cite{redlich-cleymans}. 

In \cite{redlich-canonic} the centrality dependence of the strange baryon
yields was studied based on the above concepts. The authors  
found, that the centrality
dependence of the $\Omega$ enhancement should flatten out, once the volume over
which strangeness is conserved exceeds that of about 20 time the volume of a
nucleon. Therefore, if a flat centrality dependence of the $\Omega$
enhancement is observed, one can conclude that strangeness has "percolated" at
least over a volume 20 times as large as in a nucleon-nucleon collision. This
would be a necessary but not sufficient condition for the existence of
matter. Unfortunately the results reported by the NA57 
collaboration \cite{manzari} show a steep increase of the $\Omega$ enhancement
up to the highest centralities. 

However, even if the centrality dependence of the $\Omega$-enhancement is not
completely understood, the fact that there {\em is} an $\Omega$-enhancement
clearly shows that a nucleus-nucleus collision is more than simply a
superposition of nucleon-nucleon collisions. 
And there is evidence from other observables that a certain amount of
re-scattering is taking place in heavy ion collisions. 
Flow, radial or elliptic, would be difficult to understand without
re-scattering on the partonic/hadronic level.
To which extent they are sufficient to form  matter in the Boltzmann sense
is, however, not clear. 

So have we formed matter in these collisions? A definitive answer to this
question requires additional measurements such as multi-particle correlations
of conserved quantities. At lower energy ($1-2 \,\rm AGeV$) collisions, the
measurement of kaon pairs for example provides a sensitive measurement on the
degree of equilibrium reached \cite{jeon}. At higher energies one might think
about multiple $\Omega$ production, in order to really probe the boundaries of
phase-space. 

\begin{figure}[htb]
\epsfxsize=0.6 \textwidth
 \centerline{\epsfbox{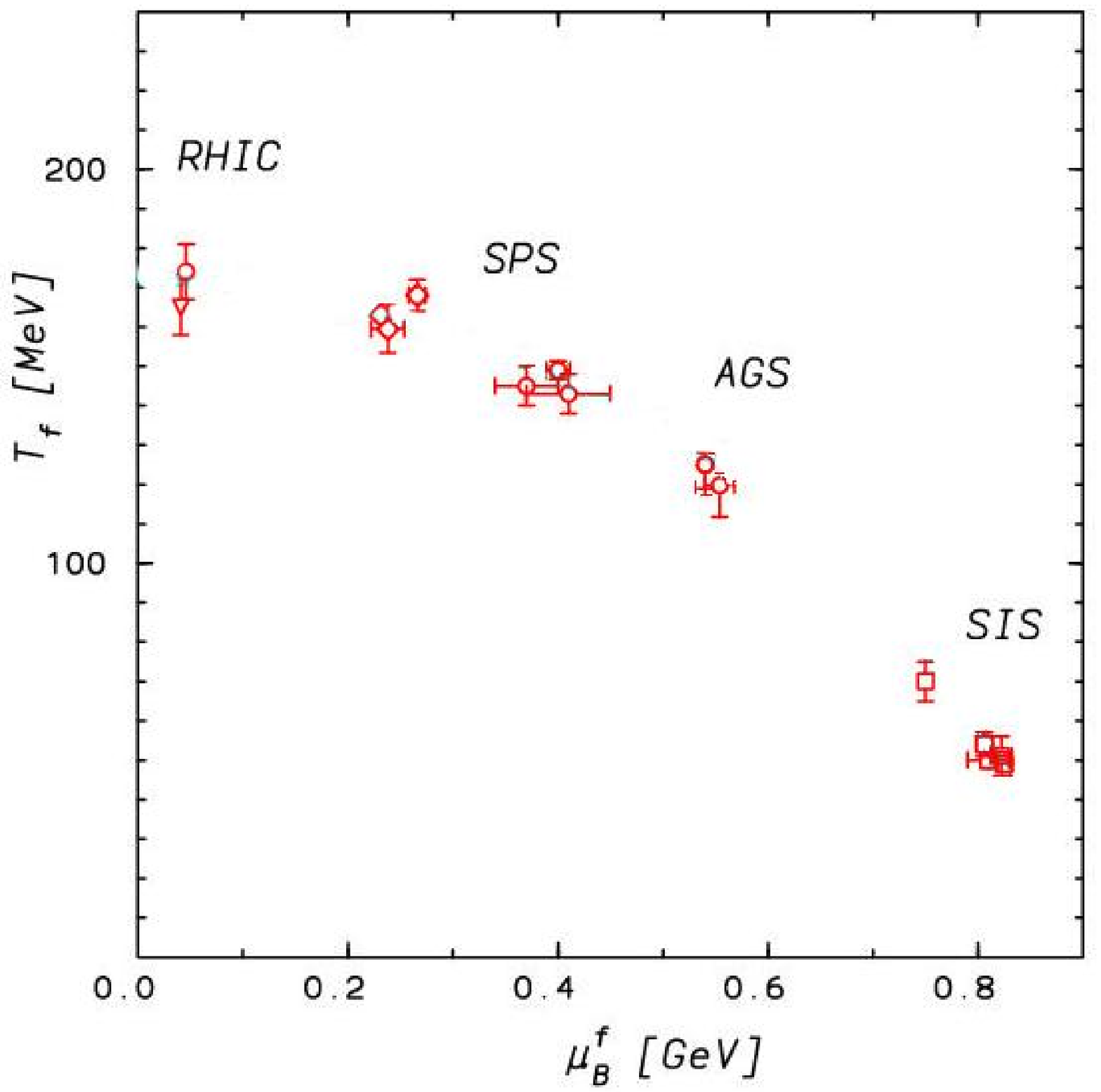}}   
 \caption{Results of thermal fits to particle ratios for different beam
 energies. Figure adapted from  \cite{redlich-cleymans}.}
\label{fig:4}
\end{figure}

But we also have no evidence {\em against} the hypothesis of thermal
equilibrium. Quite to the contrary, there is evidence for the necessary
re-scattering from flow and dilepton production as well as 
$\Omega$-enhancement.
Therefore, let us assume that we indeed have been able to create matter in
these collision. In this case, we may interpret the Lagrange multipliers $T$
and $\mu$ as temperature and chemical potential. The result of fits to system
at different collision energies \cite{pbm3,redlich-cleymans} 
is shown in Fig.\ref{fig:4}.

Does Fig.\ref{fig:4} reflect a measurement of the phase-separation line
in the QCD phase diagram? Certainly not! All it shows are the thermal
parameters at which the systems fall out of chemical equilibrium under the
assumption of unchanged particle properties. Does it tell
us about a limiting temperature? Maybe! Suppose that LHC experiments lead to
the same temperature of $T\simeq 170 \, \rm MeV$. If at the same time radial
flow increases considerably above the values observed at RHIC, then we can
conclude that indeed much more energy has been deposited into the initial 
partonic system than reflected by the final temperature. Otherwise, one could
argue that the constant temperature simply reflects the decreasing efficiency
of depositing energy in the central rapidity region. Actually the radial flow
from RHIC seems to be slightly larger than that extracted at the SPS 
\cite{NU,Phenix}.

\section{CONCLUSIONS}
We have discussed the phase-space dominance assumption in the context of
particle production in nucleon-nucleon, $e^+e^-$ and nucleus-nucleus
collisions. The fact that the statistical model is able to explain observed
particle ratios in these experiments may simply be a result of this
assumption. We also have attempted to assess the difference between
nucleus-nucleus and nucleon-nucleon collisions, and to which extent matter is
produced in the former. While there is evidence for re-scattering processes to
take place, we have not yet definitively established that  a thermal system
has been created in these collisions. A detailed study
of multi-particle correlations of conserved quanta is one possible way to
address this issue. Finally, we have argued that even if we consider the
parameters extracted from the fits to the statistical model as temperature
and chemical potential, the energy dependence of these parameters is {\em not}
a measurement of the phase separation line of QCD.

\section{ACKNOWLEDGMENTS}
I would like to thank F. Becattini, P. Braun-Munzinger, J. Knoll, A. Majumder,
J. Randrup, K. Redlich and S. Soff for useful discussions. 
This work was supported by GSI Darmstadt, and the Director,
Office of Science, Office of High Energy and Nuclear Physics,
Division of Nuclear Physics, and by the Office of Basic Energy
Sciences, Division of Nuclear Sciences, of the U.S. Department of Energy
under Contract No. DE-AC03-76SF00098.


\begin{thebibliography}{99}
\bibitem{pbm2}Figure is most recent update of result published in
  \cite{pbm}. Courtesy of P. Braun-Munzinger.
\end{thebibliography}
\end{document}